\documentclass[preprints,article,accept,pdftex,moreauthors]{Definitions/mdpi} 
\firstpage{1} 
\pubvolume{1}
\issuenum{1}
\articlenumber{0}
\pubyear{2024}
\copyrightyear{2024}
\datereceived{} 
\dateaccepted{} 
\datepublished{} 
\hreflink{https://doi.org/} 

\usepackage[
separate-uncertainty=true,
per-mode=symbol-or-fraction,
]{siunitx}
\usepackage{graphicx}
\usepackage{subcaption}
\usepackage{makecell}

\Title{TCAD Simulation of Two Photon Absorption -- Transient Current Technique measurements on Silicon Detectors and LGADs}

\TitleCitation{TCAD Simulation of the Two Photon Absorption -- Transient Current Technique}

\Author{Sebastian Pape~$^{1,2,}$*\orcidA{},
Michael Moll~$^{1,}$*\orcidC{},
Marcos Fernández García~$^{1,3}$\orcidB{}, 
 Moritz Wiehe~$^{1}$\orcidD{}}

\AuthorNames{Sebastian Pape, Michael Moll, Marcos Fernández García,  Moritz Wiehe}

\AuthorCitation{Pape, S.; M.; Moll, Fernández García, M.; Wiehe, W.}

\address{%
$^{1}$ \quad CERN, Esplanade des Particules 1, 1217 Meyrin, Switzerland\\
$^{2}$ \quad TU Dortmund University, Department of Physics - AG Kröninger, 44227 Dortmund, Germany\\
$^{3}$ \quad Instituto de Física de Cantabria (CSIC-UC), Avenida de los Castros, E-39005 Santander, Spain}

\corres{Correspondence: sebastian.pape@cern.ch, michael.moll@cern.ch}

\abstract{
Device simulation plays a crucial role in complementing experimental device
characterisation by enabling deeper understanding of internal physical
processes. However, for simulations to be trusted, experimental validation
is essential to confirm the accuracy of the conclusions drawn.
In the framework of semiconductor detector characterisation, one powerful
tool for such validation is the Two Photon Absorption - Transient Current
Technique (TPA-TCT), which allows for highly precise, three-dimensional
spatially-resolved characterisation of semiconductor detectors.
In this work, the TCAD framework Synopsys Sentaurus is used to simulate
depth-resolved TPA-TCT data for both p-type pad detectors 
(PINs) and Low Gain
Avalanche Detectors (LGADs). The simulated data are compared against
experimentally measured TPA-TCT results.
Through this comparison, it is demonstrated that TCAD
simulations can reproduce the TPA-TCT measurements,
providing valuable insights into the TPA-TCT itself.
Another significant outcome of this study is the successful simulation of the
gain reduction mechanism, which can be observed in LGADs with increasing densities of
excess charge carriers. This effect is demonstrated in an p-type LGAD with a
thickness of approximately 286 µm. The results confirm the ability of TCAD
to model the complex interaction between carrier dynamics and device gain.
}


\begin{document}




\section{Introduction}

Detector development strives for increasingly precise spatial and timing resolution to cope with the demanding requests of future high energy experiments~\cite{ECFA} and medical applications.
Increased device performance is often tackled by complex device design approaches with feature sizes in the micrometer-scale \cite{Garcia-Sciveres_2018}.
The ongoing research and development efforts benefit from precise device characterisation with high spatial resolution, which is offered by e.g. the Ion Beam Induced Charge (IBIC) method~\cite{IBIC, TRIBIC} or Transient Current Technique (TCT) based methods~\cite{TCT, kramberger_edge}.
Here, the Two Photon Absorption (TPA) TCT is applied for the characterisation of silicon detectors.
TPA is used to only generate excess charge carriers in a small volume around the focal point~\cite{HV-CMOS, Wiehe_paper}.
Thus, TPA-TCT achieves three-dimensional spatial resolution to investigate device properties that are accessible with TCT, like the collected charge (CC), the charge collection time, the electric field, etc.~\cite{Pape_thesis}.
Device simulation complements the device characterisation, by helping to gain a deeper understanding of the experimental results.
Multiple effects often appear simultaneously in experimental data and overlap each other, which complicates the understanding.
There, simulation helps to isolate overlapping effects, by investigating each effect separately.
Technology Computer Aided Design (TCAD) programs have proven to be cost-effective tools for the device simulation and are nowadays involved in every stage of device development: design exploration, process fabrication, and operation evaluation.
The latter two stages — fabrication and operation — require experimental validation, which can be achieved through techniques such as
TPA-TCT measurements.
However, using TPA-TCT measurements to benchmark TCAD simulation requires great knowledge about the measurements. 
Within this work, TCAD simulation of TPA-TCT measurements in silicon pad detectors are performed, in order to investigate the TPA-TCT itself and to evaluate the TPA-TCT as a tool to benchmark the simulation.
Firstly, an overview about the device under test (DUT) and the used TPA-TCT setup is introduced.
Then, the simulation framework is discussed and used to simulate depth resolved TPA-TCT data.
Subsequently, the experimental and simulation results are compared, with the focus on the current transients and further features relevant to the device characterisation.
In this framework, a Low Gain Avalanche Detector (LGAD) is as well simulated, where the gain reduction mechanism is correctly reproduced in the simulations and discussed. 
Finally, we summarise our work and evaluate the potential of combining the TPA-TCT measurement technique with profound TCAD simulations.

\section{Experimental methods}
\label{sec:exp_methods}
\subsection{Experimental setup}
\begin{figure}
	\centering
	\includegraphics[width=\textwidth]{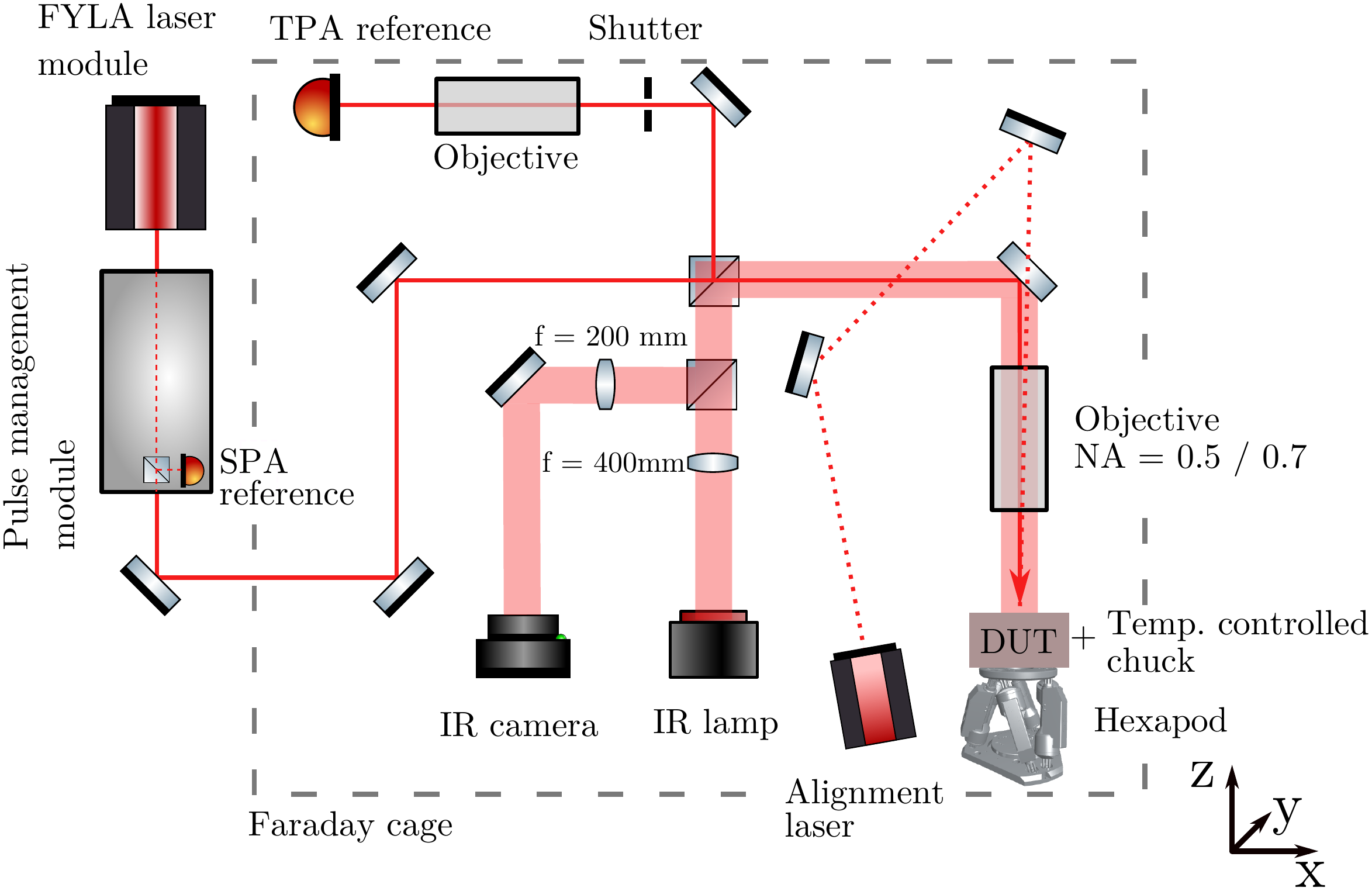}
	\caption{Schematic layout of the used TPA-TCT setup~\cite{Pape_thesis}.}
	\label{fig:setup}
\end{figure}

The measurements were performed at CERN-SSD table-top TPA-TCT setup.
A detailed description of the setup can be found in ~\cite{Wiehe_thesis, Pape_thesis}, while within this work only a brief overview 
of the setup is given.
Figure~\ref{fig:setup} shows the setup's schematic layout. 
A fibre laser that provides a central wavelength of~$\SI{1550}{\nano\meter}$, a temporal pulse width of~$\SI{430}{\femto\second}$, and a maximum pulse energy of~$\SI{10}{\nano\joule}$ is used.
For the measurements, the laser repetition rate is set to~$\SI{1}{\kilo\hertz}$ and the laser energy is adjusted, using a Neutral Density Filter (NDF), to generate a charge of ~$\SI{48.6(4)}{\femto\coulomb}$ in silicon, if not mentioned otherwise.
The laser beam can be described by a Gaussian beam with the beam waist~$w_0 = \SI{1.23}{\micro\meter}$ and the Rayleigh length~$z_{\mathrm{R}} = \SI{10.73}{\micro\meter}$.
All measurements are performed at~$\SI{20}{\celsius}$, while the DUT's ambient is flushed with dry air to establish a relative humidity of about~$\SI{0}{\percent}$.
The data acquisition is performed by an Agilent DSO9254 oscilloscope, which has a high bandwidth limit at~$\SI{2.5}{\giga\hertz}$ and a sampling rate 
of~$\SI{20}{\giga S / \second}\mathrel{\widehat{=}}\SI{50}{\pico\second/pt}$.
The average current transient of 256 single acquisitions is recorded.
A CIVIDEC C2HV transimpedance amplifier is used, which has a bandwidth from~$\SI{10}{\kilo\hertz}$ to~$\SI{2}{\giga\hertz}$ and provides an amplification by a fcator of~$\num{205.6(19)}$ ($\approx\SI{46}{\deci\bel}$)~\cite{cividec}.

\subsection{Device under test}
\label{subsec:devices}
The here presented data is acquired on a CNM~\cite{CNM} planar detector of run 8622.
The device is from the 5$^{\mathrm{th}}$ wafer and labelled J6-2 PIN detector.
Its active thickness is~$\SI{286}{\micro\meter}$ and its implantation follows a n$^+$/p/p$^+$-design.
The bulk doping concentration is extracted by capacitance-voltage measurements to about~$\SI{5e11}{\centi\meter^{-3}}$ and the device reaches full depletion at a reverse bias of~$\SI{32}{\volt}$.
Besides pad detectors, the same wafer contains Low Gain Avalanche Detectors (LGADs)~\cite{PELLEGRINI201412}.
Within this work the LGAD D8-2 is investigated.
The device is n$^+$/p$^+$/p/p$^+$-doped, its gain layer depletion occurs at~$\SI{32}{\volt}$, and it is fully depleted at~$\SI{72}{\volt}$.
The device layout is the same for both, the pad and the LGAD detector. They have an active area of about~$\SI{10.716}{\milli\meter^2}$ and a circular opening window with a diameter of about~$\SI{2.3}{\milli\meter}$.
The here presented laser scans are performed in the central position of the opening window.

\section{Simulation framework}
\label{sec:simulation_framework}
The TCAD simulations were performed with the Synopsys Sentaurus TCAD platform \cite{Sentaurus}. 3D simulations were performed for the laser pulse induced transients by exploiting the cylindrical symmetry of the problem that effectively reduces the calculations to a 2D problem with much reduced computing time. 
The sensor and in particular the LGAD doping profile are implemented following the procedure described in~\cite{Impactionization}. Input parameters for the implementation are building heavily on Capacitance - Voltage measurements of the sensors under test. The impact ionization was modelled with the parameters given in~\cite{Impactionization}.
The TPA generated charge cloud, with charge density $n_{TPA}(z,r)$, was introduced as the square of the irradiance of a Gaussian beam profile. 
\begin{align}
    n_{TPA}(z,r) = \frac{n_0}{\omega^4\left(z-z_{\mathrm{R}}\right)}
    \exp\left(-\frac{4r^2}{\omega^2\left(z-z_{\mathrm{R}}\right)} \right)
    \qquad\text{, with}\qquad \omega(z) = \omega_0 \sqrt{1+\left(\frac{z}{z_R}\right)^2}
\end{align}
with $z$ and $r$ being the cylinder coordinates, $z_{\mathrm{R}}$ the position of the focal plane, $\omega_0$ the beam waist, $z_{\mathrm{R}}$ the Rayleigh length, and $n_0$ a parameter used to scale the generated charge to the experimentally set laser intensity. The tuning of $n_0$ was performed by comparing a simulation and a measurement with the focal plane of the laser beam set to the middle of the device in depth, such that the full TPA charge cloud is confined in the detector's sensitive volume. All presented simulations of depths scans were performed with a single parameter set ($\omega_0,z_{\mathrm{R}},n_0$), shifting only the focal plane $z_R$. 
To account for the electronic shaping of the signal by the detector capacitance and the amplifier, the simulated current transient was shaped with an RC-filter.
A time constant of $\tau=\SI{700}{\pico\second}$ is used to cover for the DUT and stray impedance. The transient after filtering is convoluted with the transfer function of the amplifier that was obtained experimentally with a network analyser.

\section{Comparison between simulation and experiment}
The following sections show simulations and measurements
on the PIN sensors
for a bias voltage of~$\SI{150}{\volt}$ and a temperature of~$\SI{20}{\celsius}$.
TPA-TCT transients were simulated and measured at different device depths. Simulations were performed shifting the focal plane ($z_R$) in $\SI{10}{\micro\meter}$ steps between $\SI{-200}{\micro\meter}$ and $\SI{500}{\micro\meter}$ with respect to the top side of the DUT. 
For the measurements a stage step size along the depth of $\SI{0.5}{\micro\meter}$ was used for a very fine granularity along the device depth.

\subsection{Current transients}
\label{subsec:WFs}

\begin{figure}
	\centering
	\begin{subfigure}{0.48\textwidth}
		\centering
		\includegraphics[width=\textwidth]{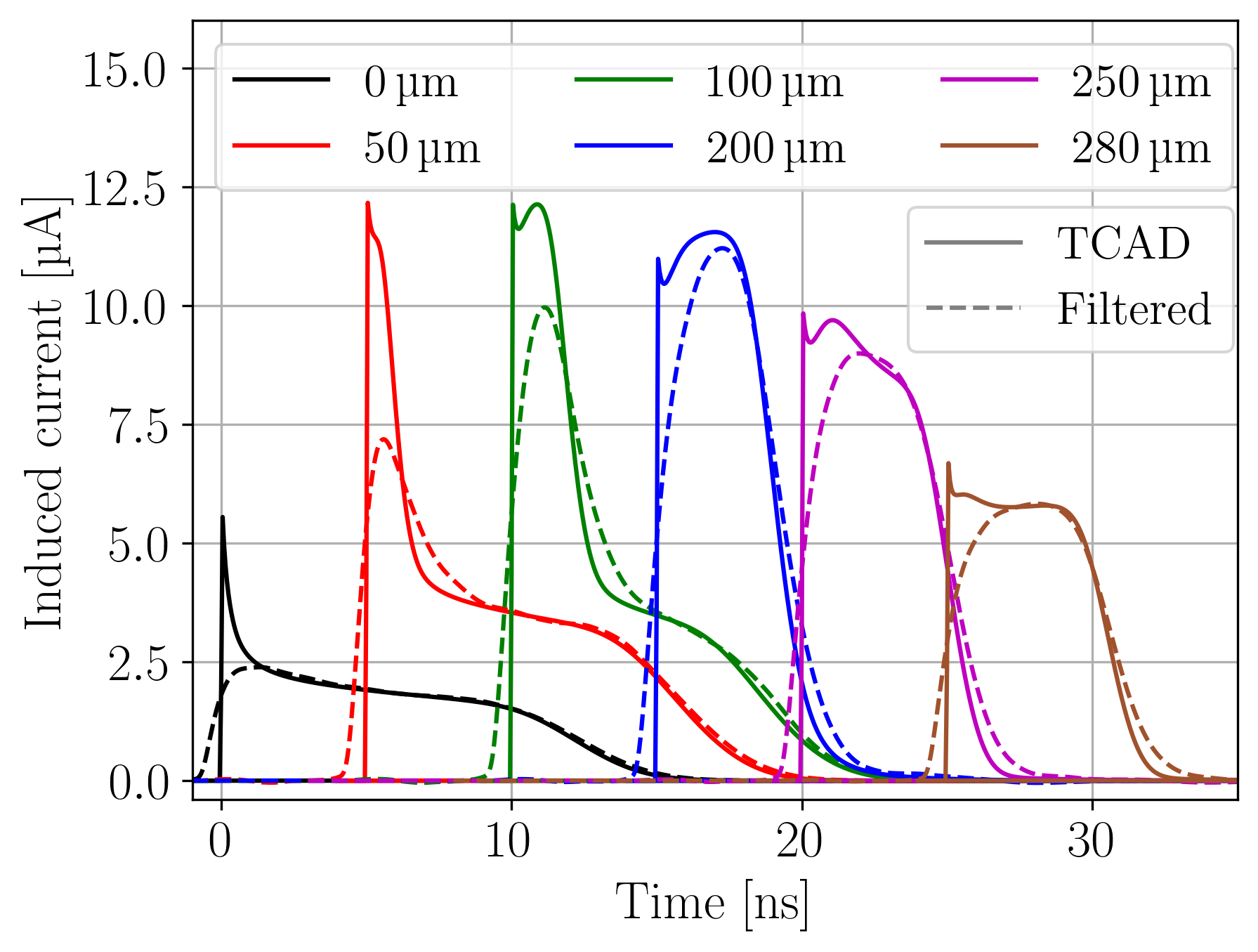}
		\caption{}
		\label{fig:WFs_TCAD}
	\end{subfigure}
	\begin{subfigure}{0.48\textwidth}
		\centering
		\includegraphics[width=\textwidth]{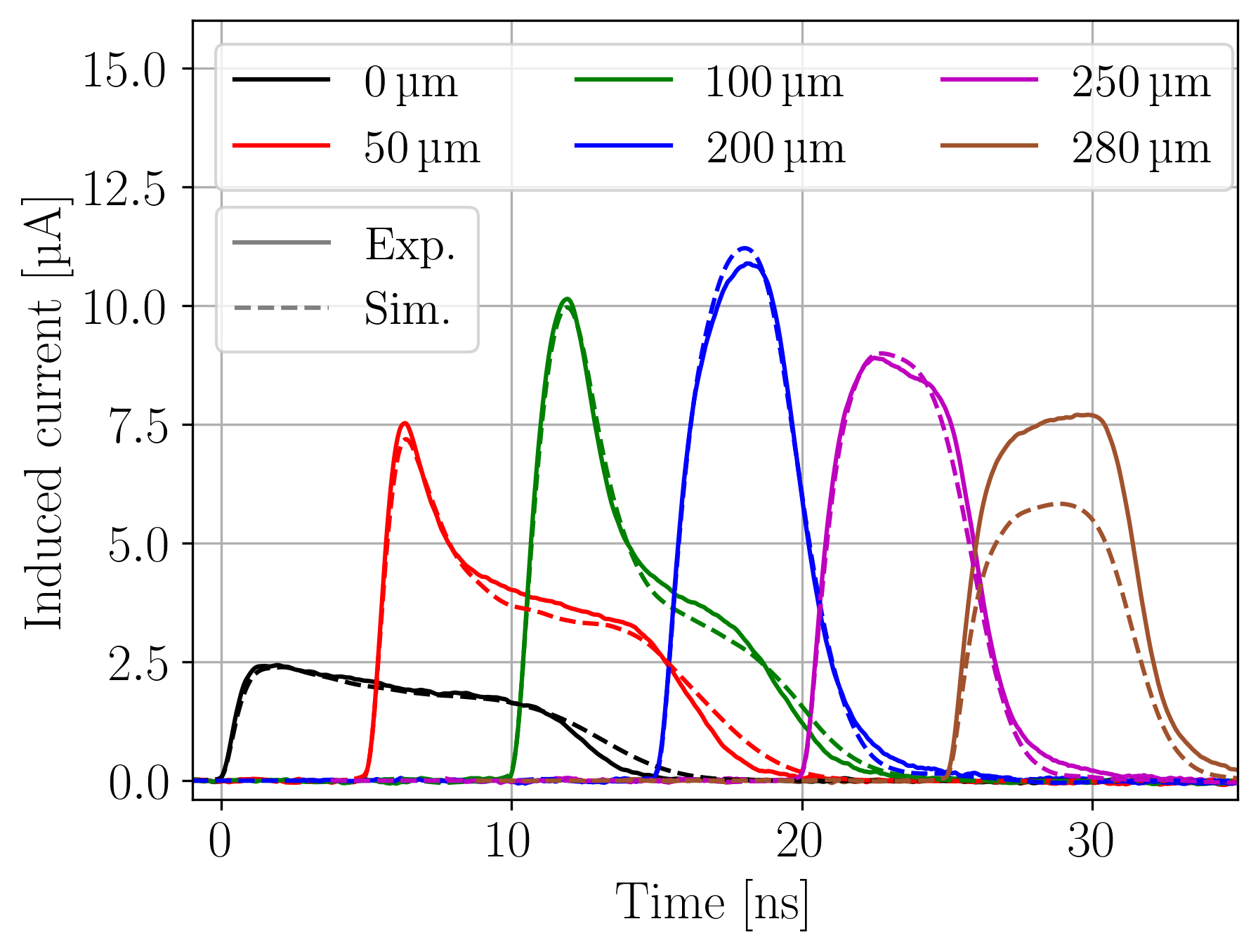}
		\caption{}
		\label{fig:WFs_comparison}
	\end{subfigure}
	\caption{(a) Simulated current transients from
    TCAD simulations (TCAD, solid lines) and after a mathematical filtering considering the readout electronics (Filtered, dashed lines). (b) Comparison between the experimental 
    (Exp., solid lines) and simulated (Sim., dashed lines) current transients. The simulated current transients contain the effects of the readout electronics and correspond to the filtered data in figure (a). The experimental current transient recoded at $z_{\mathrm{Si}}=\SI{280}{\micro\meter}$ has a significantly higher amplitude, because reflection at the sensor backside is not taken into account in the simulation.}
	\label{fig:WFs}
\end{figure}
Figure~\ref{fig:WFs_TCAD} shows 
simulated current transients of different focal depth as obtained from the TCAD simulations (TCAD) and after a subsequent mathematical filtering process (Filtered) described in Sec.~\ref{sec:simulation_framework}. 
The latter simulation step is taking into account effects from the sensor capacitance and the electronic readout. 
For a better visibility of the current transients, the starting time is increased by $\SI{5}{\nano\second}$ for each depth and the depth is given with respect to the DUT's top side. 
It can be seen that the electronic readout affects the rising edge of the simulated current transients, by rounding out the starting peak.
This is especially notable for current transients obtained close to the top side, because the electrons are collected at the top side and quickly collected, which results in a fast signal that is especially affected by the shaping.
However, the amount of collected charge (CC) is not affected by the electronic readout. An agreement within $\SI{0.1}{\percent}$ between both current transients is found.
A comparison between the \emph{filtered} current transients and the experimentally measured current transients of different depths is presented in figure~\ref{fig:WFs_comparison}.
The simulated current transients are in good agreement with the measured ones.
For the experimental current transient recorded at $z_{\mathrm{Si}}=\SI{280}{\micro\meter}$ a much higher amplitude is found compared to the simulation.
This is an effect of laser reflection at the back side silicon-air interface that leads to an increase of the laser intensity, i.e. excess charge carrier density, which is not considered in the simulation.
The simulated current transients close to the top side have an about $\SI{1}{\nano\second}$ longer collection time, while the agreement is better for current transients simulated closer to the back side.
Current transients from the top side of the DUT are dominated by the drift of holes.
Holes have an about three times smaller mobility than electrons, which make them more sensitive to uncertainties in the device and operational parameters, like the active thickness, temperature or the charge carrier drift model.
The origin of the slight disagreement might be multifold and was not further investigated.
For the following sections, the filtered simulated current transients are used for the comparison to the experimental data.

\subsection{Collected charge}
\label{subsec:collected_charge}
\begin{figure}
	\centering
	\begin{subfigure}{0.48\textwidth}
		\centering
		\includegraphics[width=\textwidth]{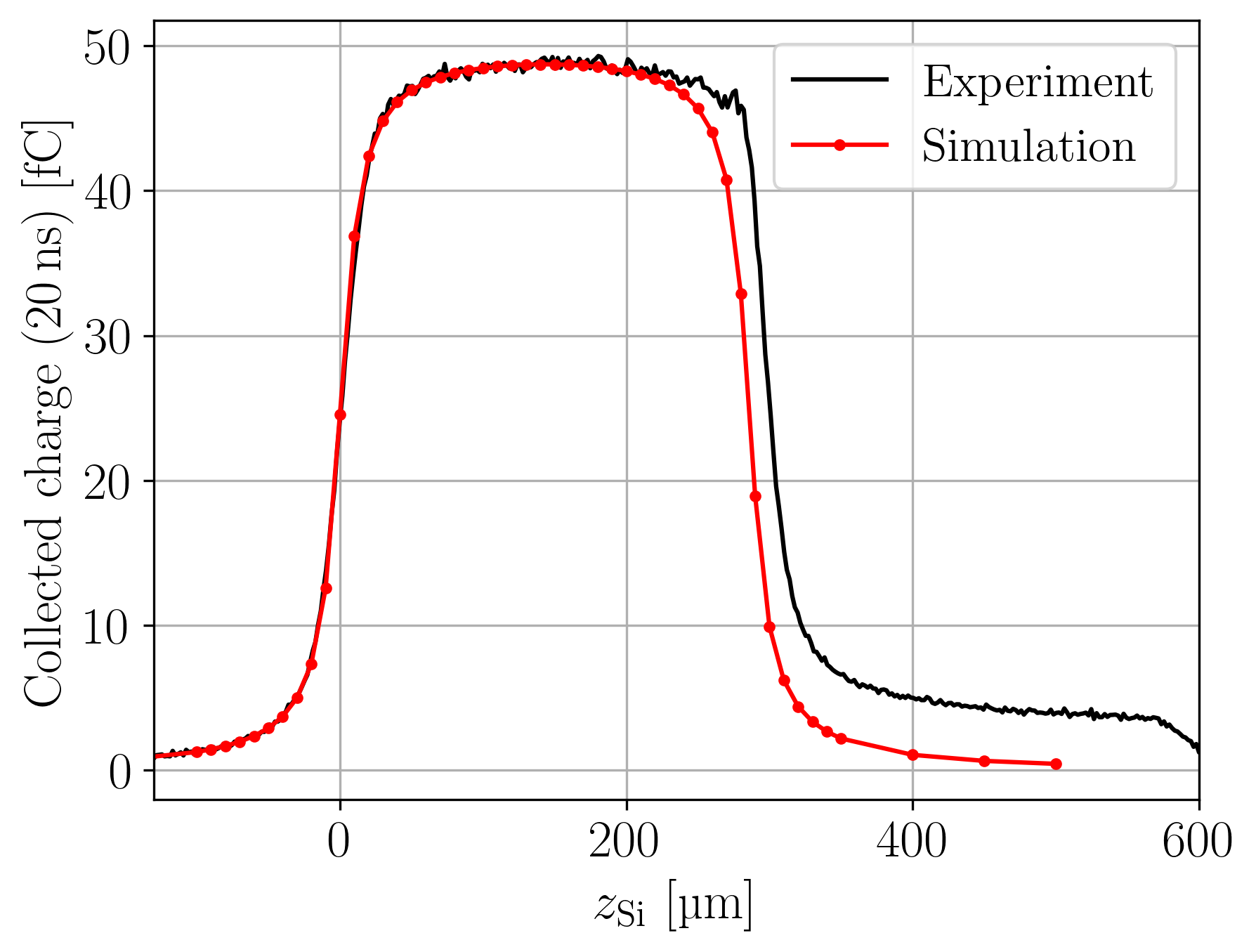}
		\caption{}
		\label{fig:cc_comparison}
	\end{subfigure}
	\begin{subfigure}{0.48\textwidth}
		\centering
		\includegraphics[width=\textwidth]{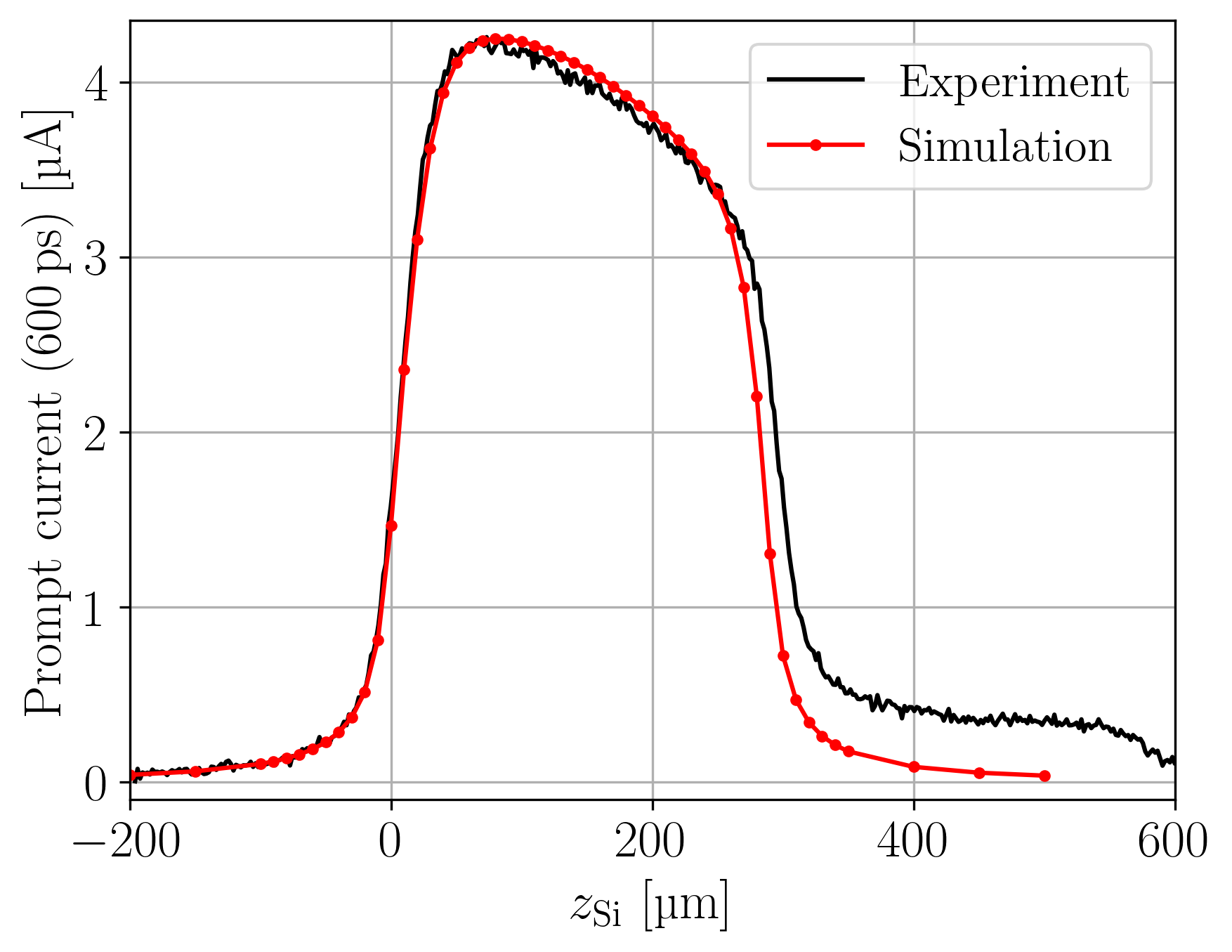}
		\caption{}
		\label{fig:pc_comparison}
	\end{subfigure}
	\caption{Simulated and measured charge collection within $\SI{20}{\nano\second}$ (a) and prompt current at $\SI{600}{\pico\second}$ (b) along the device depth.}
	\label{fig:cc_pc_comparison}
\end{figure}

Integrating the current transients over time yields the CC.
Figure~\ref{fig:cc_comparison} shows the simulated and measured CC as a function of the DUT depth, i.e. the position of the focal plane relative to the sensor surface at $z_{Si}=0$.
Both charge collection profiles show an excellent agreement at the rising edge and the middle of the detector.
At the back side, the measured CC is much higher than the simulated, because of reflection at the silicon-air interface.
Reflection is neglected within the simulation.
The reflection overshadows the falling edge in the measured CC and leads to a tail with the length of about the device thickness.
In the reflection region, approximately $\SI{10}{\percent}$ of excess charge is generated and collected with respect to the non-reflection region.
Besides the neglected reflection, an excellent agreement between the simulated and measured CC is found, which demonstrates that the depth resolution of the TPA-TCT can be simulated using TCAD.

The prompt current (PC) is a quantity to extract an approximate for the drift velocity of silicon detectors from current transient~\cite{kramberger_edge}.
It examines the current transients at a short time after light injection, here $\SI{600}{\pico\second}$, to draw conclusions on the electric field at the position of excess charge generation.
Figure~\ref{fig:pc_comparison} shows the simulated and measured prompt current.
The higher measured prompt current at the back side is again an effect of the reflection that is not considered in the simulation.
Simulation and experiment are in good agreement, which gives confidence in the simulated model of the DUT.

\subsection{Time over threshold}
\label{subsec:tot}

\begin{figure}
	\centering
	\begin{subfigure}{0.48\textwidth}
		\centering
		\includegraphics[width=\textwidth]{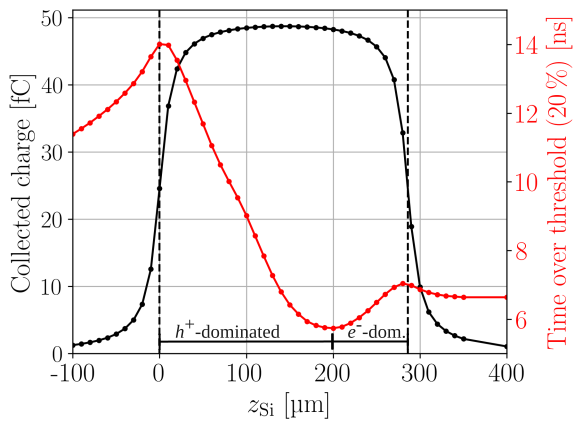}
		\caption{}
		\label{fig:cc_tcoll_sim}
	\end{subfigure}
	\begin{subfigure}{0.48\textwidth}
		\centering
		\includegraphics[width=\textwidth]{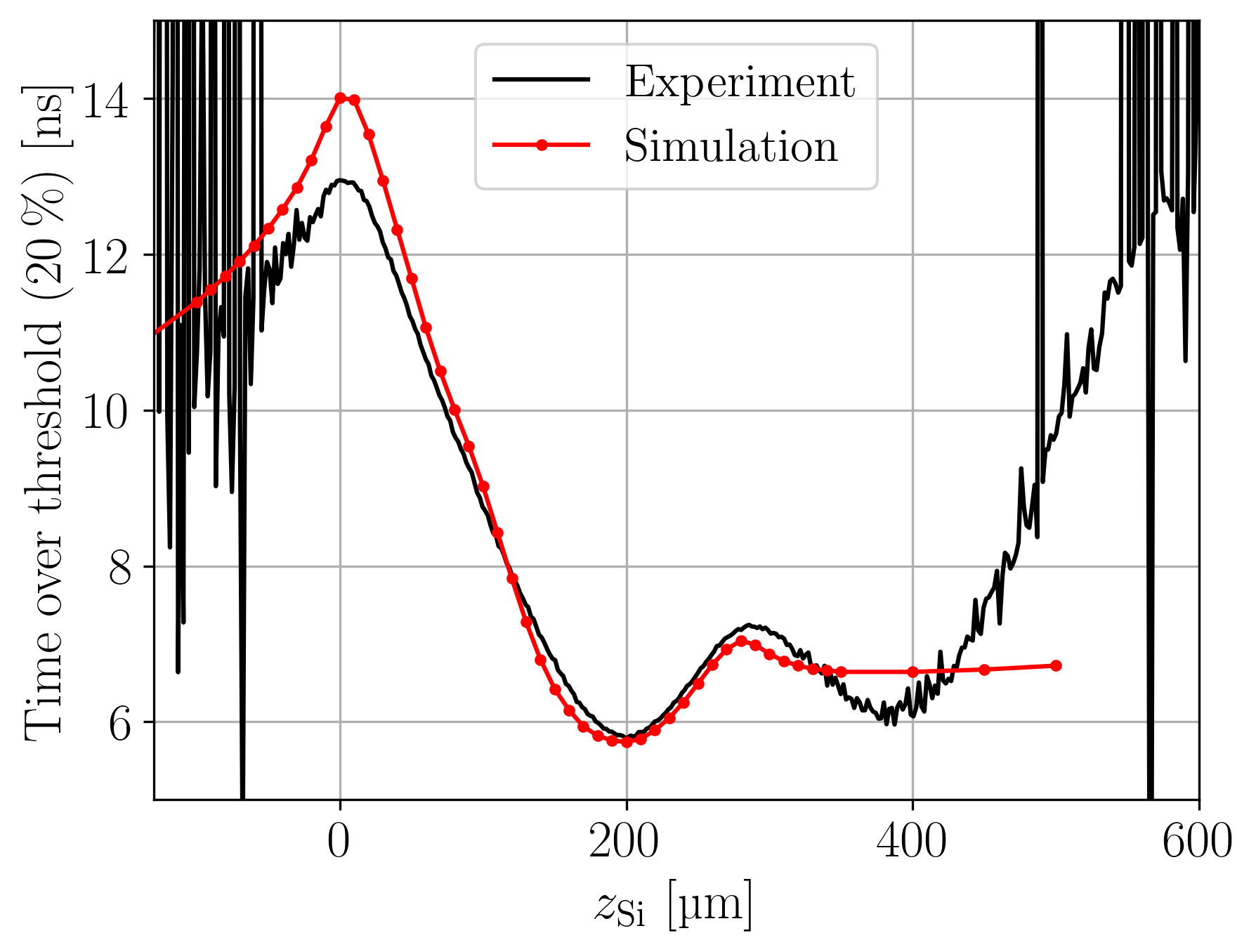}
		\caption{}
		\label{fig:tcoll_tpc_sim}
	\end{subfigure}
	\caption{(a) Simulated collected charge profile overlapped with the simulated time over threshold profile along the device depth. The regions where holes or electrons dominate the collection time in the current transients are indicated. (b) Comparison between the simulated and measured time over threshold profile along the device depth. The experimental time over threshold profile is affected by reflection, which mirrors the region~$\SI{0}{\micro\meter} < z_{\mathrm{Si}} < \SI{286}{\micro\meter}$ at the back side to $z_{\mathrm{Si}} > \SI{286}{\micro\meter}$.}
	\label{fig:tcoll}
\end{figure}

The time over threshold (ToT) is here defined as the time the current transient is above a given fraction of its amplitude.
The ToT can be used as a measure for the collection time of the charge.
Low amplitude fractions thresholds can result in higher noise in the experimental data, compared to too high thresholds that can result in less satisfactory conclusions on the collection time~\cite{Pape_thesis}.
Here a fraction $\SI{20}{\percent}$ of the current transient's amplitude is used for the ToT.
Figure~\ref{fig:cc_tcoll_sim} shows the simulated CC overlapped with the ToT as a function of the device depth.
Current transients from excess charge generation at or close to the top side are dominated by the contribution of holes.
Therefore, the ToT is longer compared to current transients, where electrons have the dominant contribution, because holes have only a third of the electrons mobility and thus require a longer drift time to be collected.
At about $z_{\mathrm{Si}}=\SI{200}{\micro\meter}$ the ToT has a minimum, at the position where electrons and holes have the same drift time.
The position of the minimum depends on the selected amplitude's fraction~\cite{Pape_thesis}.\\
Furthermore, a decrease of the ToT is observed when the focal point is located outside of the active volume ($z_{\mathrm{Si}} < \SI{0}{\micro\meter}$ or $z_{\mathrm{Si}} > \SI{286}{\micro\meter}$) but still close enough so that light can enter the active volume due to the extension of the focal region ($z_{\mathrm{R}}$). In principle, the position of the focal point represents the position of the dominant contribution to the induced current.
From this argumentation one would expect that the ToT should be constant after the focal point exited the active volume.
However, at the device's boundaries the median of the excess charge carrier density shifts so that the dominant contribution is no longer at the focal point's position.
The median moves back inside the active volume and the ToT decreases.\\
The ToT for a charge generation directly at the top or back side could be found by a linear fit that extrapolates the linear trend in the bulk towards the boundaries.
Following this procedure yields an ToT of about $\SI{14.5}{\nano\second}$ at the top and about $\SI{7.4}{\nano\second}$ at the back side, compared to the $\SI{14}{\nano\second}$ and $\SI{7}{\nano\second}$ extracted from the current transients, respectively.
Hence, data recorded at the device's boundaries needs to be taken with care, because the median excess charge, i.e. the dominant contribution to the current transient, does not coincide with focal point position.
Meaning that the TPA-TCT can yield less satisfactory sampling at the device boundaries.
This effect is described in more detail in the appendix of reference~\cite{Pape_thesis}.
The effect does not affect the CC, as it becomes only apparent in intensity independent quantities that depend on the temporal shape of the current transient.

Figure~\ref{fig:tcoll_tpc_sim} presents the comparison between the simulated and the measured ToT.
Both ToT profiles show the same features and the agreement in absolute values is best at the middle and back side of the DUT.
A disagreement of about $\SI{1}{\nano\second}$ is found at the top and the reason for the disagreement could be multifold, as discussed in section~\ref{subsec:WFs}.
The back side in the measured ToT is again affected by the reflection, which leads in this case to a mirroring of the focal point, i.e. the excess charge distribution, and thus leads to a peak at the back side~\cite{Pape_2022}.
However, the features are present in both ToT profiles and the before discussed shift of the median at the top side is experimentally found as well.
Applying the linear fitting procedure in order to compensate for the shifting median to the experimental ToT at the top side yields a ToT of $\SI{13.6}{\nano\second}$, compared to the $\SI{13}{\nano\second}$ extracted from the measured current transients.
The fitting procedure can not be applied to the experimental back side ToT, because it is dominated by reflection.
Anyhow the influence of the shifting median is expected to be smaller when the focal point is reflected back into the active volume, as such reflection folds back part of the excess charge carrier volume, which reduces the influence of a shifting median.

\subsection{Weighted prompt current}

\begin{figure}[bh]
	\centering
    \begin{subfigure}{0.48\textwidth}
		\centering
		\includegraphics[width=\textwidth]{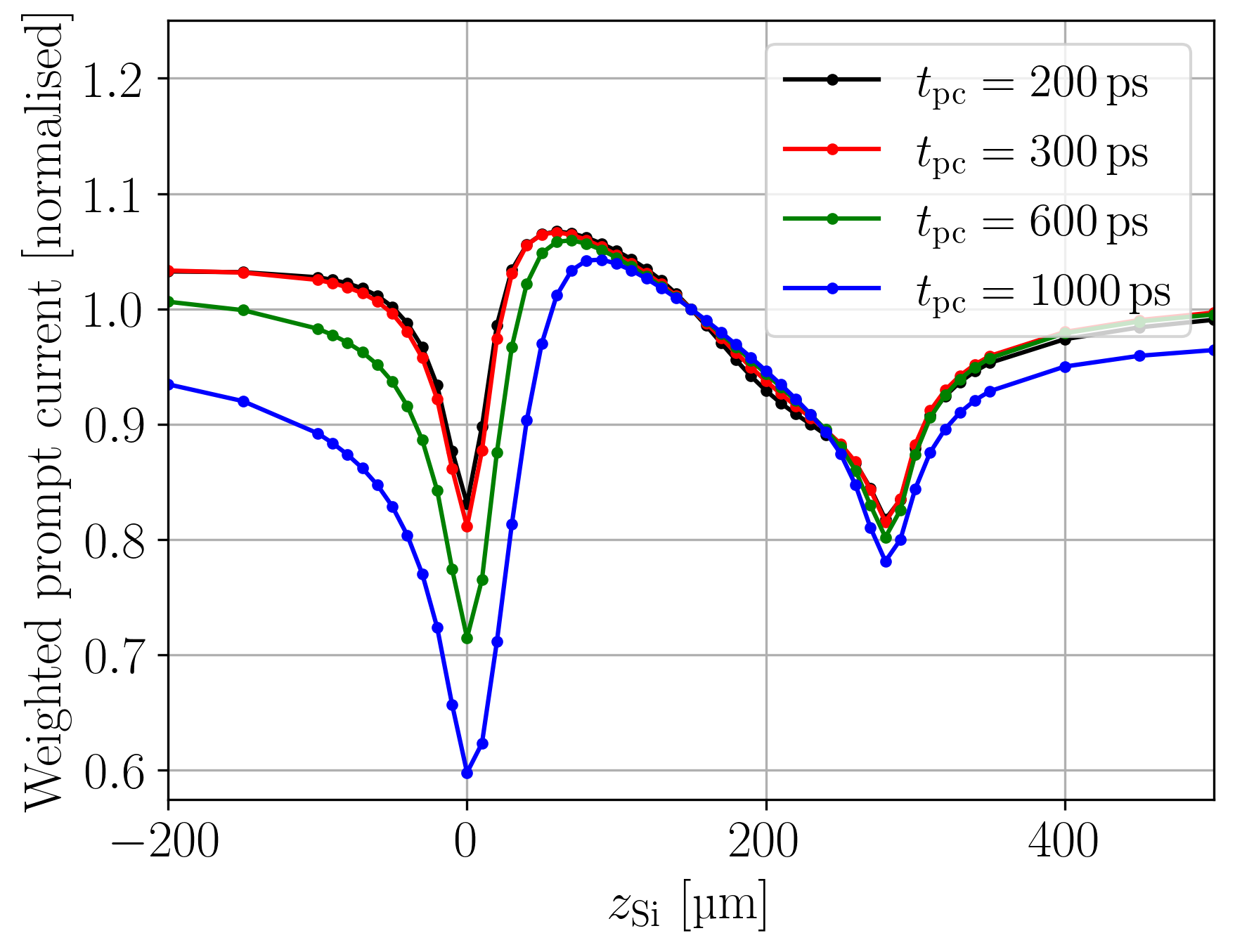}
		\caption{}
		\label{fig:wpc_tpc_sim}
	\end{subfigure}
	\begin{subfigure}{0.48\textwidth}
		\centering
		\includegraphics[width=\textwidth]{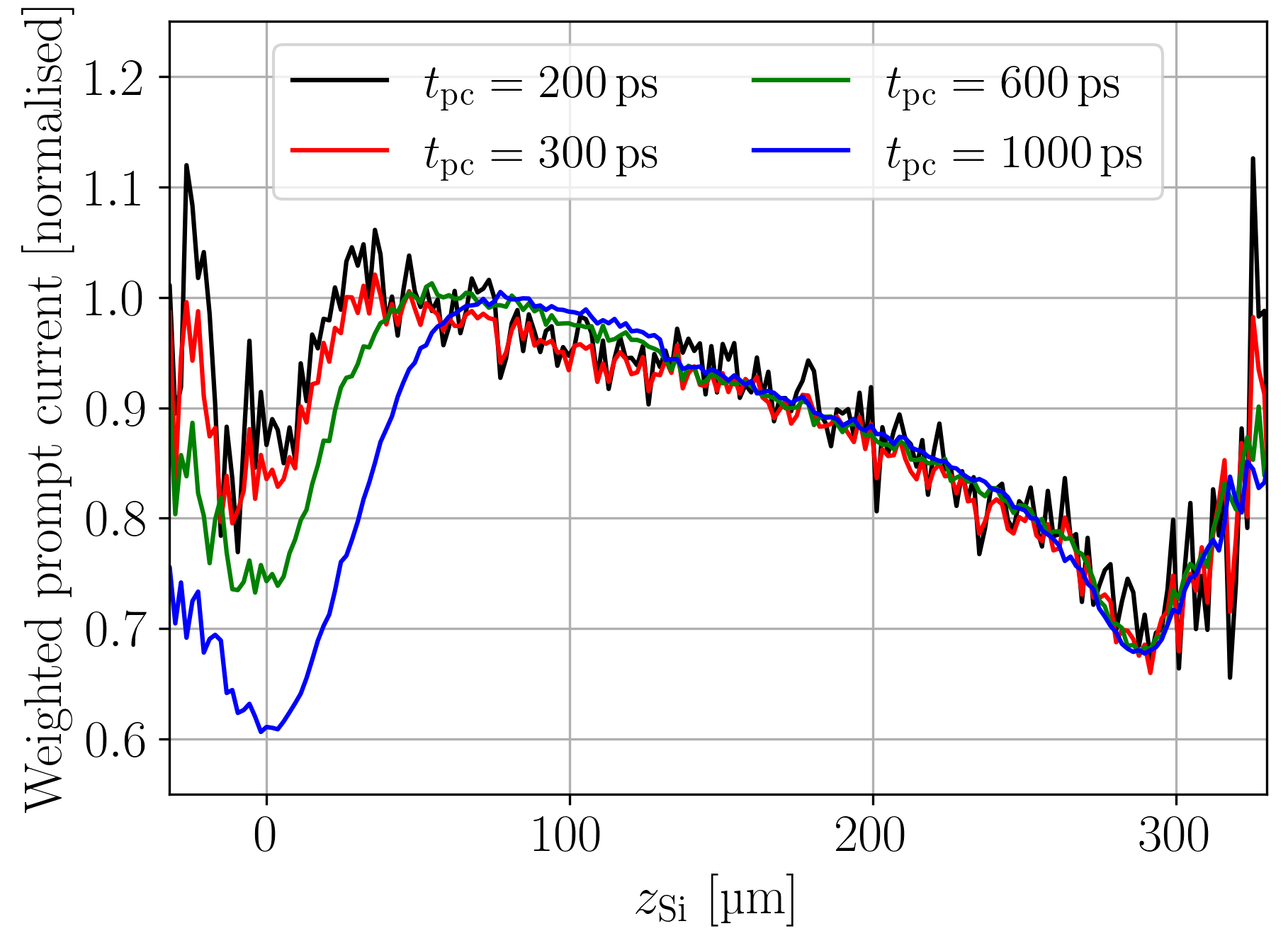}
		\caption{}
		\label{fig:wpc_tpc_exp}
	\end{subfigure}
	\caption{Simulated (a) and measured (b) weighted prompt current for different prompt current times along the  device depth. The valleys at the device boundaries are an effect of a shifting median of the excess charge (see text in section~\ref{subsec:tot}).}
	\label{fig:wpc_tpc}
\end{figure}
The weighted prompt current (WPC) method was introduced in order to correct for a position dependent charge carrier generation~\cite{Pape_wpc}.
Similar to the prompt current, it is an approximation to the drift velocity and weighting is obtained by dividing the prompt current by the CC.
Figure~\ref{fig:wpc_tpc} shows the simulated (a) and measured (b) WPC for different prompt current times $t_{\mathrm{pc}}$ along the DUT depth.
The absolute values of the simulated and measured WPC profiles are in good agreement, as the absolute values of the prompt current and CC are in good agreement.
Absolute values for the WPC profiles are not presented here, because all WPC profiles are normalised to the value at $z_{Si} = \SI{150}{\micro\meter}$, to ease the comparability between the different $t_{\mathrm{pc}}$ values.
The simulated and measured WPC profiles show the same features.
Valleys are found at the top and back side and the depth of the valleys at the top side increase with increasing $t_{\mathrm{pc}}$.
The valleys are caused by charge collection within $t_{\mathrm{pc}}$.
The longer $t_{\mathrm{pc}}$ is selected the more excess charge carriers are collected and the less prompt current is extracted at $t_{\mathrm{pc}}$, which leads to a decreasing WPC.
Increasing bias voltages exacerbate this effect as the charge is than even faster collected.
Hence, the $t_{\mathrm{pc}}$ should be selected as small as possible, while maintaining sufficient signal-to-noise ratio (SNR) and obeying the maximum possible probing band width.
Band width limitations of the readout electronics are the reason why the valley depth is not significantly changing between $t_{\mathrm{pc}}=\SI{200}{\pico\second}$ and $t_{\mathrm{pc}}=\SI{300}{\pico\second}$.
The appearance of the valleys indicates that the WPC needs to be taken with care, when it is evaluated at the device's boundaries.
Further, the symmetry around the top and back side interface are caused by the shifting median of the excess charge distribution, as discussed for the ToT in section~\ref{subsec:tot}.

\section{Gain reduction on a Low Gain Avalanche Diodes}
To further validate the TPA-TCT TCAD simulation approach, the complex problem of gain reduction in Low Gain Avalanche Diodes (LGADs) was targeted. LGADs are sensors with an intrinsic signal amplification that is originating from a moderate signal gain through impact ionization in an embedded high field region called gain layer. Deposition of excess charge carriers in the sensor with subsequent multiplication in the gain layer can lead to a high concentration of free charges in the gain layer. The electric field produced by these charges can reduce the high electric field strength in the gain layer and consequently suppress further impact ionization processes and thus the gain. Such effect on LGADs has been first observed by comparing measurements with a beta source to pulsed laser measurements \cite{Curras-2022} and later evaluated in more detail in experiments using ion beams \cite{IBIC-LGAD}, and pulsed lasers \cite{PAPE2022167190} for the charge generation. 

The complexity in simulating this effect is lying in the fact that the charge generated in the avalanche process is impacting on the electric field in the gain layer and thus, for the simulation process, has a feedback-loop to the impact ionization process itself. In such situation, the drifting charges and the electric field cannot be treated independently, which prevents the use of fast signal solvers that are based on static electric fields, i.e. solving the poisson equation to acquire the electric field prior to the simulation of the movement of the excess charges forming the sensor signal. A simulation solving the coupled semiconductor equations, like the present TCAD simulation, is required, as well as a very good knowledge of the gain layer doping profiles. 

The purpose of this work is to demonstrate that TCAD simulations can reproduce the gain suppression effect. More profound studies aiming to develop LGAD structures with lower gain suppression or to produce parametrizations that could be used to model the gain suppression in fast signal solvers with static electric fields are not addressed here. 

\begin{figure}[bth]
	\centering
    \begin{subfigure}{0.48\textwidth}
		\centering
		\includegraphics[width=\textwidth]{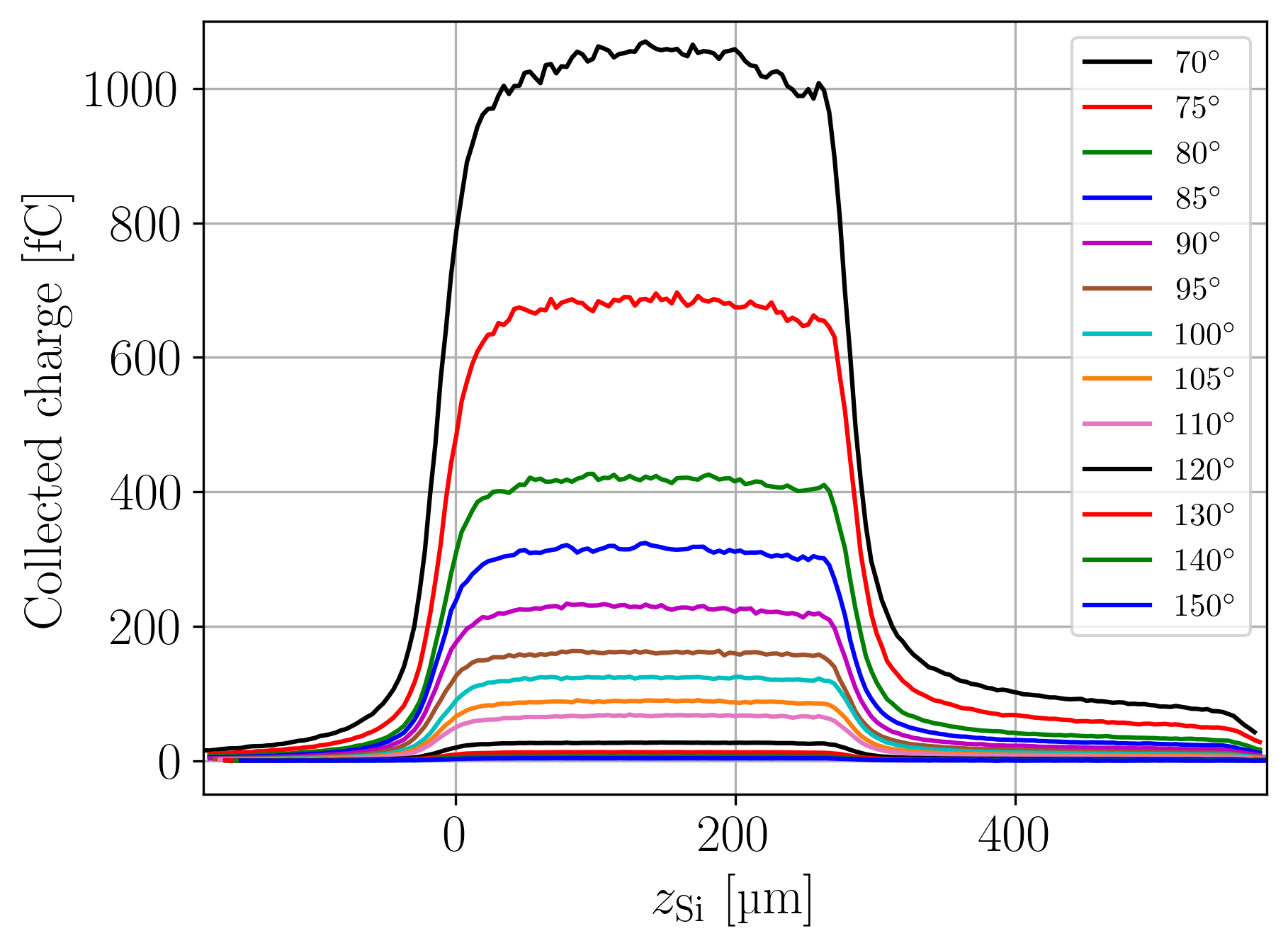}
		\caption{}		\label{fig:chargeprofile_pin_vs_lgad}
	\end{subfigure}
	\begin{subfigure}{0.48\textwidth}
		\centering
		\includegraphics[width=\textwidth]{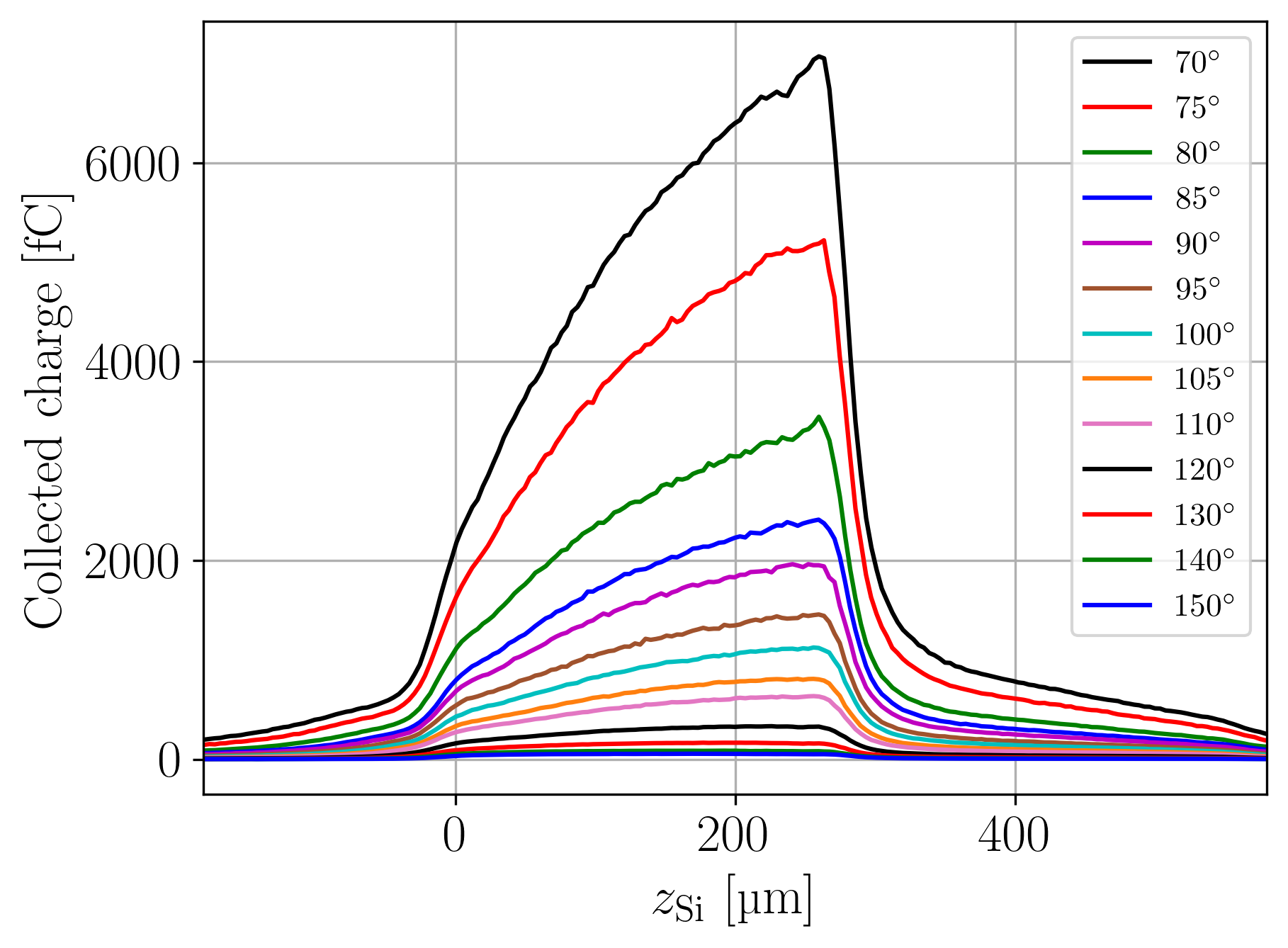}
		\caption{}
		\label{fig:chargeprofile_pin_vs_lgad}
	\end{subfigure}
	\caption{Experimental data: Charge collection within $\SI{20}{\nano\second}$ as function of the device depth for a PIN diode (a) and an LGAD (b). Both devices are fully depleted and operated at~$\SI{-860}{\volt}$ (PIN) and~$\SI{-900}{\volt}$ (LGAD), respectively. The charge collection profiles were obtained with different laser pulse intensities. The legend gives the angular setting of the NDF with the highest laser pulse intensity for the smallest angular setting.
 }
	\label{fig:chargeprofile_pin_vs_lgad}
\end{figure}

Figure~\ref{fig:chargeprofile_pin_vs_lgad} demonstrates the gain suppression effect in comparing TPA-TCT CC measurements performed on a PIN diode (Figure~\ref{fig:chargeprofile_pin_vs_lgad}a) and on an LGAD structure (Figure~\ref{fig:chargeprofile_pin_vs_lgad}b). The TPA-TCT experimental settings and the used devices are those described in section \ref{sec:exp_methods} and the measurement method is equivalent to the one described in section \ref{subsec:collected_charge} for the PIN diode. Several CC profiles for different laser pulse energies are shown in the figure, where the legend gives the angular setting of the Neutral Density Filter (NDF) with the highest laser pulse intensity obtained for the smallest angular setting. 

The figure shows that the CC profile shape for the PIN does not depend on the laser pulse intensity and that for the PIN the same charge is collected as function of deposition depth. For the LGAD the situation is different due to the internal gain of the device. A constant charge collection profile is only seen for very small laser pulse intensities. With rising intensity, less and less charge multiplication is observed for charge deposition close to the top surface at $z_{Si}=0$ (i.e. close to the gain layer) of the device compared to charge deposition at the backside (i.e. far away from the gain layer). The characteristic shape of the CC profile gave rise to the naming as shark-fin effect. It is explained by the fact that electrons of the charge clouds deposited closer to the backside of the device have to drift a longer distance to the gain layer than charge clouds deposited closer to the front side~\cite{PAPE2022167190}. Due to diffusion, the electron charge clouds from the back side of the device are further expanded than the ones deposited closer to the front side upon arrival at the gain layer. Correspondingly, the charge density is lower for the clouds arriving from the back of the device and thus less gain suppression is observed. 
\begin{figure}[h]
	\centering
    \begin{subfigure}{0.48\textwidth}
		\centering
		\includegraphics[width=\textwidth]{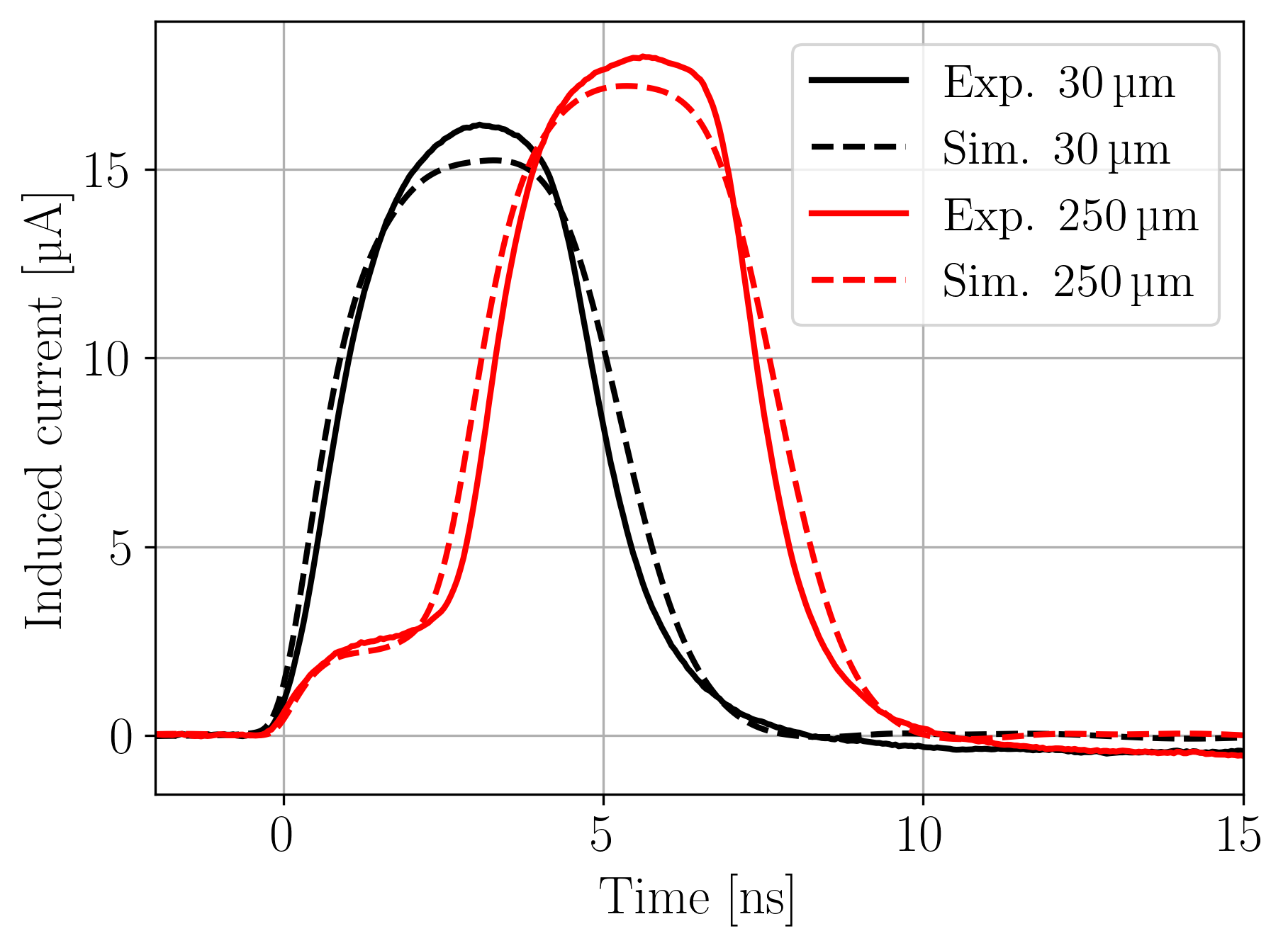}
		\caption{}		\label{fig:lgad_waveforms}
	\end{subfigure}
	\begin{subfigure}{0.48\textwidth}
		\centering
		\includegraphics[width=\textwidth]{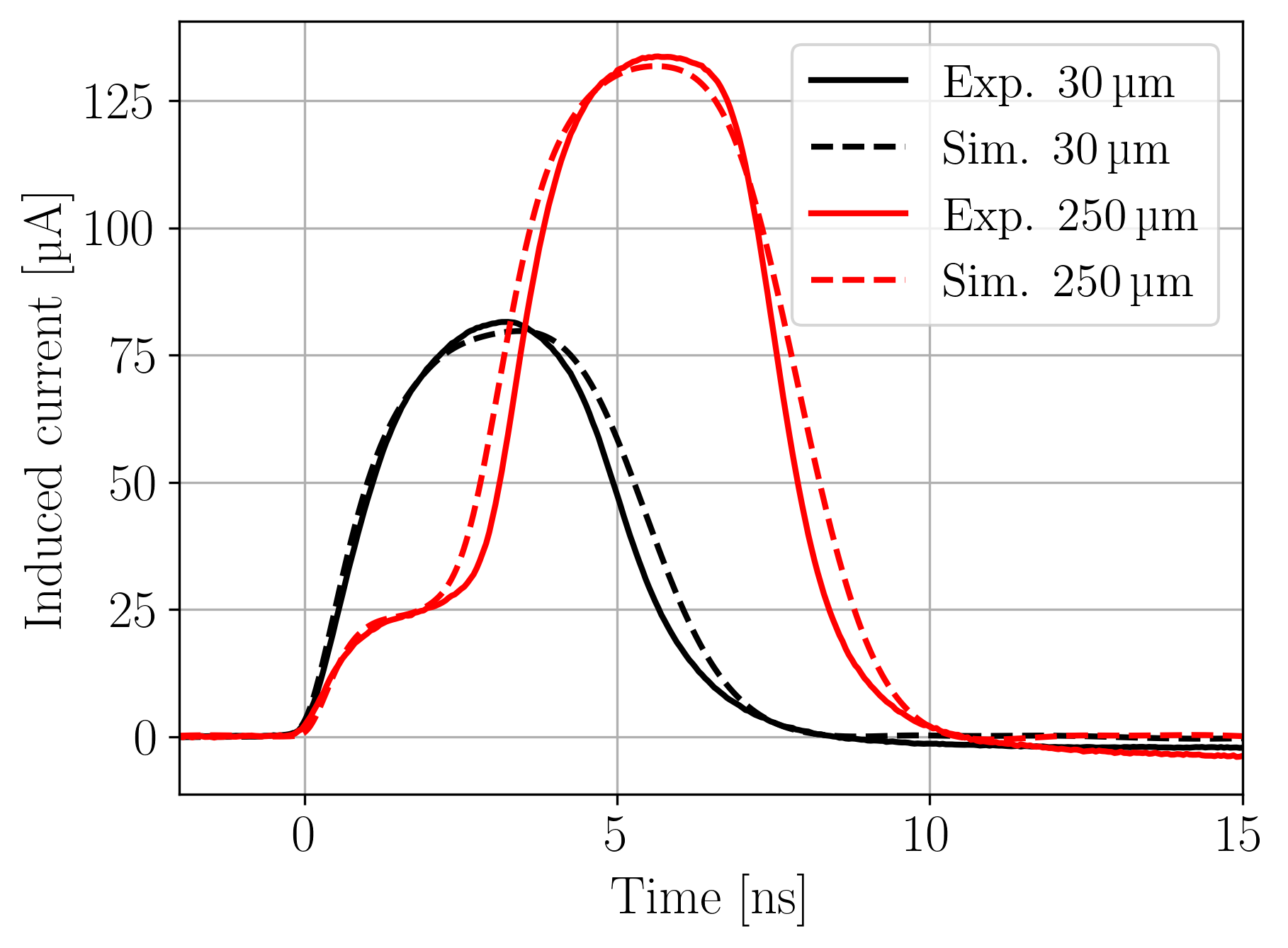}
		\caption{}
		\label{fig:lgad_waveforms}
	\end{subfigure}
	\caption{Waveforms recorded at the top side ($\SI{30}{\micro\meter}$) and back side ($\SI{250}{\micro\meter}$) of the  LGAD for an injected charge of (a)~$\SI{6.5}{\femto\coulomb}$ (NDF $\SI{140}{\degree}$) and (b)~$\SI{68}{\femto\coulomb}$  (NDF $\SI{110}{\degree}$). The solid lines represent the experimental data and the dashed lines the corresponding TCAD simulations.
 }
	\label{fig:lgad_waveforms}
\end{figure}

Figure~\ref{fig:lgad_waveforms} shows waveforms recorded with charge injection close to the top of the LGAD sensor at $\SI{30}{\micro\meter}$ and close to the back of the sensor at $\SI{250}{\micro\meter}$ for two different laser pulse intensities resulting in (a)~$\SI{6.5}{\femto\coulomb}$ and (b)~$\SI{68}{\femto\coulomb}$ of deposited charge in the sensor. It is clearly visible that for the ten times higher charge deposition, a more significant difference between the CC (i.e. integral of the waveform) at the front and at the back is observed. In addition to the experimental data, the simulated waveforms are shown. A good agreement is seen between simulation and experiment, taking into account that the simulated data is the unaltered simulation result, without any additional scaling factor. 
\begin{figure}[h]
	\centering
    \begin{subfigure}{0.48\textwidth}
		\centering
		\includegraphics[width=\textwidth]{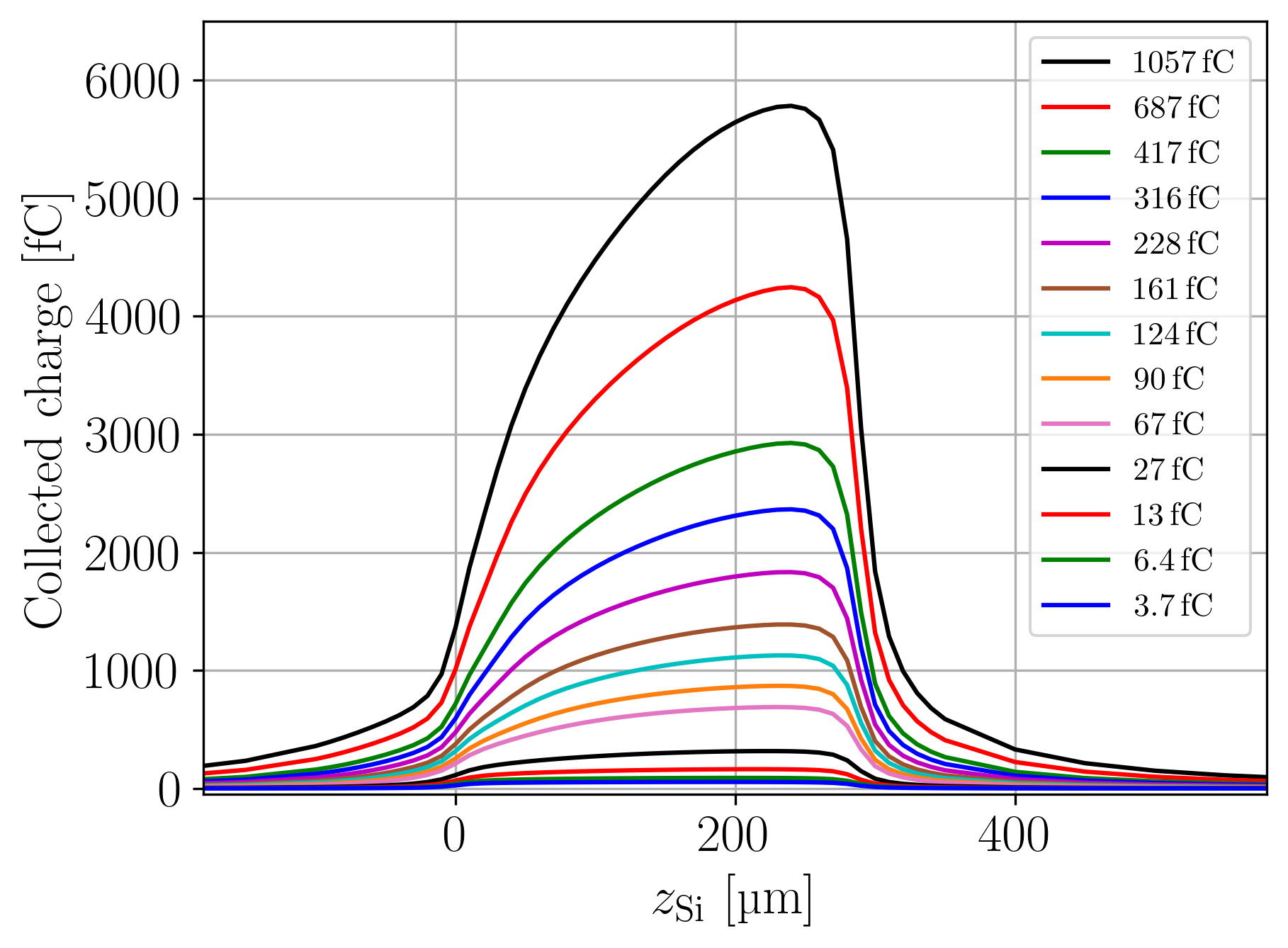}
		\caption{}		\label{fig:TCAD_LGAD_chargeprofile}
	\end{subfigure}
	\begin{subfigure}{0.48\textwidth}
		\centering
		\includegraphics[width=\textwidth]{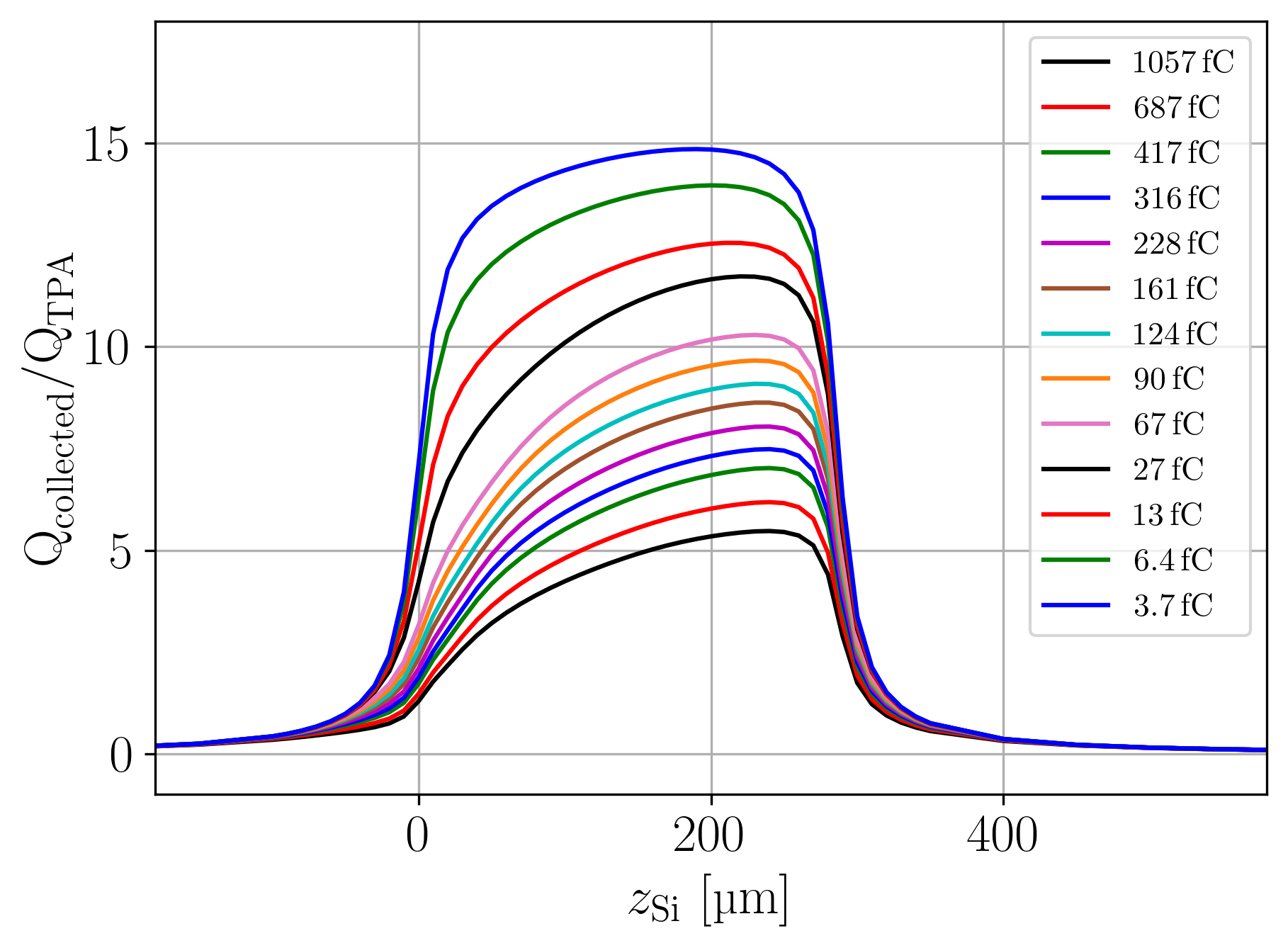}
		\caption{}
		\label{fig:TCAD_LGAD_gainprofile}
	\end{subfigure}
	\caption{Simulation: (a) Charge collection within $\SI{20}{\nano\second}$ along the device depth for the LGAD with the same laser intensity settings as the experimental data shown in Fig.\ref{fig:chargeprofile_pin_vs_lgad}. (b) The gain as function of depth in the device as obtained by dividing the CC in the LGAD by the CC in the centre of a PIN for the same laser intensity . 
} 
	\label{fig:TCAD_LGAD_chargeprofile}
\end{figure}

Integration of the simulated waveforms results in one value for the CC as function of focal depth and intensity of the laser pulse. The CC profiles for the various simulated laser intensities are given in Figure~\ref{fig:TCAD_LGAD_chargeprofile}a. 
The shark-fin effect observed for the experimental data (Figure~\ref{fig:chargeprofile_pin_vs_lgad}) is correctly reproduced in the TCAD simulations. Some deviations towards the backside of the LGAD are visible, but are known to be originating from the reflection of the laser beam at the back side metal of the device that was not taken into account in the simulation.
The effect was already describe in section~\ref{subsec:collected_charge} for the PIN detector data shown in Figure~\ref{fig:cc_comparison}.

Dividing the CC by the charge $Q_{TPA}$, defined as the charge injected when the laser is focused close to the middle of the sensor bulk ($\approx \SI{150}{\micro\meter}$), results in the data presented in Figure\ref{fig:TCAD_LGAD_gainprofile}, which can be seen as a gain profile. For the lowest injected charge, the highest gain of about 15 is observed, with only small differences between injection at the front side 
and the back side. With further rising laser pulse intensity (rising injected charge) the gain is suppressed and a more pronounced difference between the gain obtained at the front and the back side.
\begin{figure}
	\centering
	\includegraphics[width=0.8\textwidth]{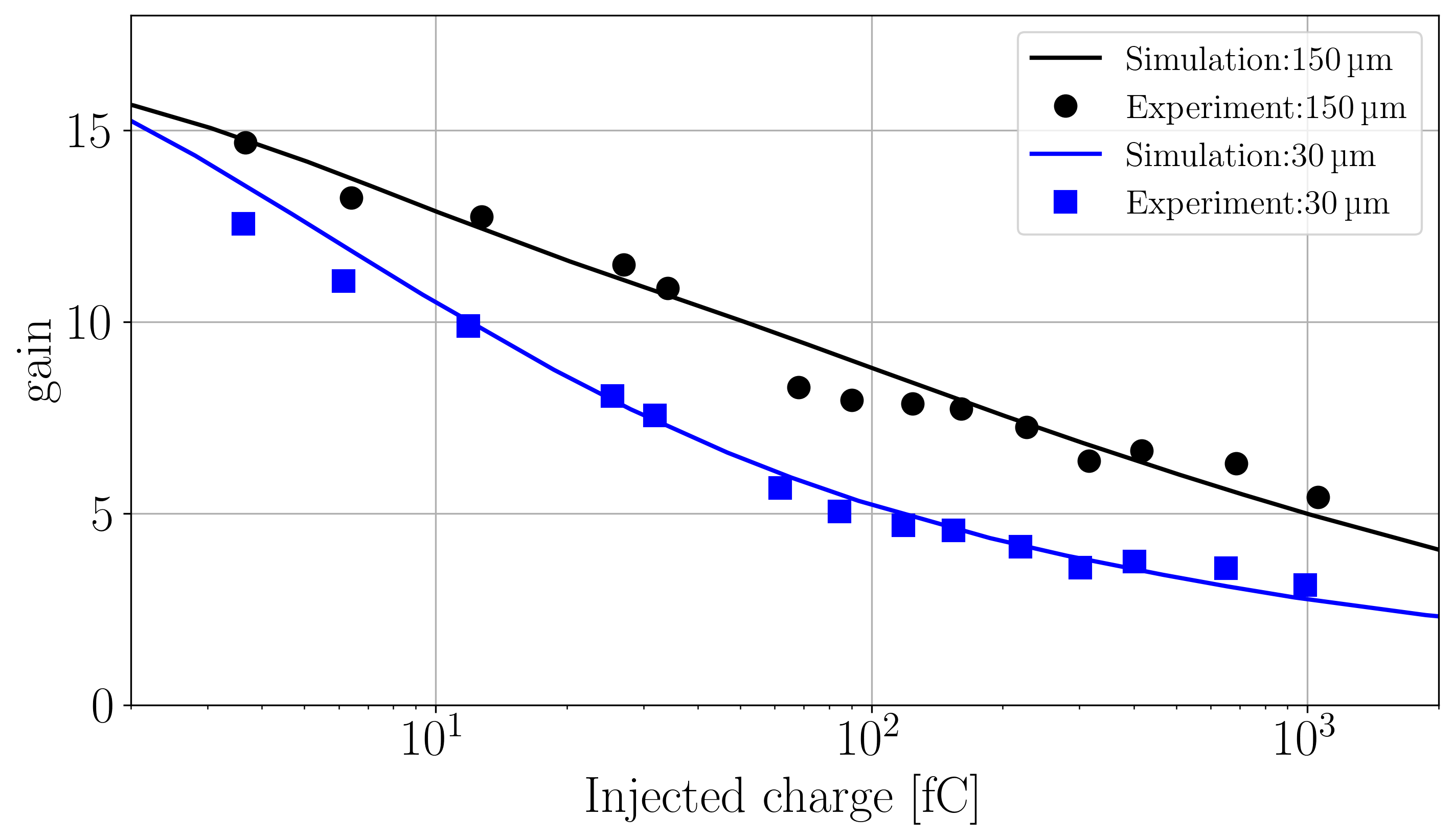}
	\caption{LGAD gain 
    at $\SI{-900}{\volt}$ as function of TPA-TCT generated charge deposited in $\SI{30}{\micro\meter}$ and $\SI{150}{\micro\meter}$ depths of the device. The lines represent the simulation data simulation and the points the experimental values.
 }
	\label{fig:gain-suppression}
\end{figure}

The gain values obtained at two different injection depths are given in Figure~\ref{fig:gain-suppression} for the experimental and the simulated data over three orders of magnitude in injected charge. The agreement between simulation and experiment clearly confirms that the shark-fin effect and the gain suppression mechanism in general, can successfully be simulated in TCAD simulations, employing a local impact-ionization model within the standard drift-diffusion simulation approach.

\section{Conclusions}

We demonstrated that TCAD device simulations can be used to simulate depth resolved TPA-TCT measurements.
A very good agreement between simulation and experiment was found. Simulation helps to interpret measurement data, as the influence of operational and device parameters can be easily distinguished.
In this work the simulation was used to interpret Time over Threshold (ToT) and Weighted Prompt Current (WPC) profiles and successfully compared to measured data.
For example, it was found that the position of the median of the TPA generated excess charge carrier distribution, i.e the position of the main contribution to the current transient, does not align with the focal point's position, when the focal point is moved beyond the interfaces of the DUT. 
Instead, the position of the main contribution to the current transient is moved back inside the active volume.
Hence, measurements at the boundaries of the TPA-TCT need to be evaluated with care.

In the second part of the paper we presented a comprehensive set of TPA-TCT experimental data on the gain suppression effect in LGAD sensors and demonstrated that the so-called shark-fin effect can be correctly reproduced in TCAD simulations. We covered the full range of experimentally accessible charge injection from $\SI{3.6}{\femto\coulomb}$ to $\SI{1000}{\femto\coulomb}$ with a highly focused laser beam with a beam waist of $\SI{1.2}{\micro\meter}$ and a Rayleigh length of $\SI{11}{\micro\meter}$. We demonstrated a good quantitative agreement between simulation and experimental data of better than $\SI{10}{\percent}$ in the collected charge (CC) for the investigated cases. 
It is thus proven that the effect can be well simulated in the framework of the drift-diffusion model with a local impact ionization model. This finding gives potential to further TCAD studies on optimising e.g. LGAD gain layers for lowering the gain suppression or towards parametrization of gain suppression effects to be implemented in fast signal simulators building on static electric fields, which are not impacted by the drifting charge.

The TPA-TCT is a powerful high precision technique with 3D resolution. In this work we demonstrated that it is also an ideal tool to benchmark TCAD device simulations and that the simulations are very useful to understand the measured data. The simulations enhances the reach of the overall technique and should proof very useful in future works to characterise and understand very complex or radiation damaged device structures in the application field of sensors or electronics.

\vspace{6pt} 



\authorcontributions{
Conceptualisation, M.M. and S.P.; 
methodology, M.F.G., M.M., S.P. and M.W.; 
software, M.F.G., M.M. and S.P.;
validation, M.M. and S.P.; 
formal analysis, M.M. and S.P.; 
investigation, M.M. and S.P.; 
resources, M.M.; 
data curation, S.P.; 
writing---original draft preparation, M.M. and S.P.; 
writing---review and editing, M.F.G., M.M. and M.W.; 
visualization, M.M. and S.P.; 
supervision, M.M.; 
project administration, M.M.; 
funding acquisition, M.M. 
All authors have read and agreed to the published version of the manuscript.
}

\funding{
This project was performed within the framework of RD50 and DRD3 and has received funding from the European Union's Horizon 2020 Research and Innovation programme under GA no 101004761 (AIDAinnova), the Wolfgang Gentner Program of the German Federal Ministry of Education and Research (grant no. 05E18CHA), and the CERN Knowledge Transfer Fund, through a grant awarded in 2017.
}


\informedconsent{Not applicable}

\dataavailability{
The data presented in this study are available on request from the corresponding author. The data are not publicly available due to the complexity of the data and to maintain an overview on usage.
} 

\acknowledgments{
Not applicable
}

\conflictsofinterest{The authors declare no conflict of interest. The funders had no role in the design of the study; in the collection, analyses, or interpretation of data; in the writing of the manuscript; or in the decision to publish the~results.} 



\abbreviations{Abbreviations}{
The following abbreviations are used in this manuscript:\\

\noindent 
\begin{tabular}{@{}ll}
CC & Collected charge\\
DUT & Device under test\\
NDF & Neutral density filter\\
SNR & Signal-to-noise ratio\\
TCAD & Technology Computer Aided Design\\
TCT & Transient Current Technique\\
ToT & Time over threshold\\
TPA & Two Photon Absorption\\
WPC & Weighted prompt current\\
\end{tabular}
}

\begin{adjustwidth}{-\extralength}{0cm}

\reftitle{References}


\bibliography{references.bib}

\begin{thebibliography}{999}

\bibitem[{ECFA Detector R\&D Roadmap Process Group}(2020)]{ECFA}
{ECFA Detector R\&D Roadmap Process Group}.
\newblock {The 2021 ECFA detector research and development roadmap}.
\newblock Technical report, Geneva,  2020.
\newblock {\url{https://doi.org/10.17181/CERN.XDPL.W2EX}}.

\bibitem[Garcia-Sciveres and Wermes(2018)]{Garcia-Sciveres_2018}
Garcia-Sciveres, M.; Wermes, N.
\newblock A review of advances in pixel detectors for experiments with high rate and radiation.
\newblock {\em Reports on Progress in Physics} {\bf 2018}, {\em 81},~066101.
\newblock {\url{https://doi.org/10.1088/1361-6633/aab064}}.

\bibitem[Manfredotti \em{et~al.}(1998)Manfredotti, Fizzotti, Polesello, Vittone, Truccato, {Lo Giudice}, Jaksic, and Rossi]{IBIC}
Manfredotti, C.; Fizzotti, F.; Polesello, P.; Vittone, E.; Truccato, M.; {Lo Giudice}, A.; Jaksic, M.; Rossi, P.
\newblock IBIC and IBIL microscopy applied to advanced semiconductor materials.
\newblock {\em Nuc. Instrum. Methods Phys. Res. B: Beam Interactions with Materials and Atoms} {\bf 1998}, {\em 136-138},~1333--1339.
\newblock Ion Beam Analysis, {\url{https://doi.org/10.1016/S0168-583X(97)00829-X}}.

\bibitem[Vizkelethy \em{et~al.}(2001)Vizkelethy, Brunett, Walsh, James, and Doyle]{TRIBIC}
Vizkelethy, G.; Brunett, B.; Walsh, D.; James, R.; Doyle, B.
\newblock Investigation of the electronic properties of cadmium zinc telluride (CZT) detectors using a nuclear microprobe.
\newblock {\em Nucl. Instrum. Methods Phys. Res. A: Accel., Spectrom., Detect. Assoc. Equip.} {\bf 2001}, {\em 458},~563--567.
\newblock Proc. 11th Inbt. Workshop on Room Temperature Semiconductor X- and Gamma-Ray Detectors and Associated Electronics, {\url{https://doi.org/10.1016/S0168-9002(00)00917-7}}.

\bibitem[Eremin \em{et~al.}(1996)Eremin, Strokan, Verbitskaya, and Li]{TCT}
Eremin, V.; Strokan, N.; Verbitskaya, E.; Li, Z.
\newblock Development of transient current and charge techniques for the measurement of effective net concentration of ionized charges ({N}{\ensuremath{_{eff}}}) in the space charge region of p-n junction detectors.
\newblock {\em Nuclear Instruments and Methods in Physics Research A: Accelerators, Spectrometers, Detectors and Associated Equipment} {\bf 1996}, {\em 372},~388--398.

\bibitem[Kramberger \em{et~al.}(2010)Kramberger, Cindro, Mandić, Mikuž, Milovanović, Zavrtanik, and Žagar]{kramberger_edge}
Kramberger, G.; Cindro, V.; Mandić, I.; Mikuž, M.; Milovanović, M.; Zavrtanik, M.; Žagar, K.
\newblock {I}nvestigation of {I}rradiated {S}ilicon {D}etectors by {E}dge-{TCT}.
\newblock {\em IEEE Transactions on Nuclear Science} {\bf 2010}, {\em 57},~2294--2302.

\bibitem[Fernández~García \em{et~al.}(2017)Fernández~García, González~Sánchez, Jaramillo~Echeverría, Moll, Montero, Moya, Palomo~Pinto, and Vila]{HV-CMOS}
Fernández~García, M.; González~Sánchez, J.; Jaramillo~Echeverría, R.; Moll, M.; Montero, R.; Moya, D.; Palomo~Pinto, F.R.; Vila, I.
\newblock On the determination of the substrate effective doping concentration of irradiated {HV-CMOS} sensors using an edge-{TCT} technique based on the {T}wo-{P}hoton-{A}bsorption process.
\newblock {\em JINST} {\bf 2017}, {\em 12},~C01038.

\bibitem[Wiehe \em{et~al.}(2021)Wiehe, Fernández~García, Moll, Montero, Palomo, Vila, Muñoz-Marco, Otgon, and Pérez-Millán]{Wiehe_paper}
Wiehe, M.; Fernández~García, M.; Moll, M.; Montero, R.; Palomo, F.R.; Vila, I.; Muñoz-Marco, H.; Otgon, V.; Pérez-Millán, P.
\newblock Development of a Tabletop Setup for the Transient Current Technique Using Two-Photon Absorption in Silicon Particle Detectors.
\newblock {\em IEEE Transactions on Nuclear Science} {\bf 2021}, {\em 68},~220--228.
\newblock {\url{https://doi.org/10.1109/TNS.2020.3044489}}.

\bibitem[Pape(2024)]{Pape_thesis}
Pape, S.
\newblock Characterisation of Silicon Detectors Using the Two Photon Absorption – Transient Current Technique.
\newblock PhD thesis, TU Dortmund University,  2024.

\bibitem[Wiehe(2021)]{Wiehe_thesis}
Wiehe, M.
\newblock {Development of a Two-Photon Absorption - TCT system and Study of Radiation Damage in Silicon Detectors}.
\newblock PhD thesis, Albert-Ludwigs-Universität Freiburg im Breisgau,  2021.

\bibitem[civ()]{cividec}
CIVIDEC Instrumentation GmbH, Vienna, Austria.
\newblock \url{https://cividec.at/index.php}, accessed on 2024-02-13.

\bibitem[CNM(2018)]{CNM}
Centre {N}acional de {M}icroelectrònica, {IMB-CNM-CSIC}, {B}arcelona, {S}pain.
\newblock \url{https://www.imb-cnm.csic.es}, accessed on 2024-02-13.

\bibitem[Pellegrini \em{et~al.}(2014)Pellegrini, Fernández-Martínez, Baselga, Fleta, Flores, Greco, Hidalgo, Mandić, Kramberger, Quirion, and Ullan]{PELLEGRINI201412}
Pellegrini, G.; Fernández-Martínez, P.; Baselga, M.; Fleta, C.; Flores, D.; Greco, V.; Hidalgo, S.; Mandić, I.; Kramberger, G.; Quirion, D.;  et~al.
\newblock Technology developments and first measurements of Low Gain Avalanche Detectors (LGAD) for high energy physics applications.
\newblock {\em Nucl. Instrum. Methods Phys. Res. A: Accel., Spectrom., Detect. Assoc. Equip.} {\bf 2014}, {\em 765},~12--16.
\newblock HSTD-9 2013 - Proceedings of the 9th International "Hiroshima" Symposium on Development and Application of Semiconductor Tracking Detectors, {\url{https://doi.org/10.1016/j.nima.2014.06.008}}.

\bibitem[Sen()]{Sentaurus}
Synopsys Sentaurus TCAD.
\newblock \url{https://www.synopsys.com/}, accessed on 2024-07-05.

\bibitem[Currás~Rivera and Moll(2023)]{Impactionization}
Currás~Rivera, E.; Moll, M.
\newblock Study of Impact Ionization Coefficients in Silicon With Low Gain Avalanche Diodes.
\newblock {\em IEEE Transactions on Electron Devices} {\bf 2023}, {\em 70},~2919--2926.
\newblock {\url{https://doi.org/10.1109/TED.2023.3267058}}.

\bibitem[Pape \em{et~al.}(2022)Pape, {Fern{\'{a}}ndez Garc{\'{\i}}a}, Moll, Montero, Palomo, Vila, and Wiehe]{Pape_2022}
Pape, S.; {Fern{\'{a}}ndez Garc{\'{\i}}a}, M.; Moll, M.; Montero, R.; Palomo, F.; Vila, I.; Wiehe, M.
\newblock Characterisation of irradiated and non-irradiated silicon sensors with a table-top two photon absorption {TCT} system.
\newblock {\em Journal of Instrumentation} {\bf 2022}, {\em 17},~C08011.
\newblock {\url{https://doi.org/10.1088/1748-0221/17/08/c08011}}.

\bibitem[Pape \em{et~al.}(2023)Pape, Currás, Fernández~García, and Moll]{Pape_wpc}
Pape, S.; Currás, E.; Fernández~García, M.; Moll, M.
\newblock Techniques for the Investigation of Segmented Sensors Using the Two Photon Absorption-Transient Current Technique.
\newblock {\em Sensors} {\bf 2023}, {\em 23}.
\newblock {\url{https://doi.org/10.3390/s23020962}}.

\bibitem[Currás \em{et~al.}(2022)Currás, Fernández, and Moll]{Curras-2022}
Currás, E.; Fernández, M.; Moll, M.
\newblock Gain reduction mechanism observed in Low Gain Avalanche Diodes.
\newblock {\em Nuclear Instruments and Methods in Physics Research Section A: Accelerators, Spectrometers, Detectors and Associated Equipment} {\bf 2022}, {\em 1031},~166530.
\newblock {\url{https://doi.org/https://doi.org/10.1016/j.nima.2022.166530}}.

\bibitem[Jiménez-Ramos \em{et~al.}(2022)Jiménez-Ramos, García~López, García~Osuna, Vila, Currás, Jaramillo, Hidalgo, and Pellegrini]{IBIC-LGAD}
Jiménez-Ramos, M.C.; García~López, J.; García~Osuna, A.; Vila, I.; Currás, E.; Jaramillo, R.; Hidalgo, S.; Pellegrini, G.
\newblock Study of Ionization Charge Density-Induced Gain Suppression in LGADs.
\newblock {\em Sensors} {\bf 2022}, {\em 22}.
\newblock {\url{https://doi.org/10.3390/s22031080}}.

\bibitem[Pape \em{et~al.}(2022)Pape, Currás, {Fernández García}, Moll, Montero, Palomo, Vila, Wiehe, and Quintana]{PAPE2022167190}
Pape, S.; Currás, E.; {Fernández García}, M.; Moll, M.; Montero, R.; Palomo, F.; Vila, I.; Wiehe, M.; Quintana, C.
\newblock First observation of the charge carrier density related gain reduction mechanism in LGADs with the Two Photon Absorption-Transient Current Technique.
\newblock {\em Nuclear Instruments and Methods in Physics Research Section A: Accelerators, Spectrometers, Detectors and Associated Equipment} {\bf 2022}, {\em 1040},~167190.
\newblock {\url{https://doi.org/https://doi.org/10.1016/j.nima.2022.167190}}.

\end{thebibliography}

\end{adjustwidth}
\end{document}